\documentstyle[preprint,aps]{revtex}
\newcommand{\rms}[2]{$#1_{\text{#2}}$}
\newcommand{\mrms}[2]{#1_{\text{#2}}}
\begin{document}
\preprint{ gr-qc/9401025 }
\draft
\title{  Geodesic structure of the 2 + 1 black hole   }
\addtocounter{footnote}{1}
\author {
 Norman Cruz$^{1,2}$\thanks{Electronic address: ncruz@usachvm1.usach.cl},
 Cristi\'an Mart\'{\i}nez$^{1}$\thanks{Electronic address:
 jzanelli@abello.seci.uchile.cl} and Leda Pe\~na$^{1}$ }
\address{
 $^1$ Departamento de F\'{\i}sica, Facultad de Ciencias,
Universidad de Chile,\\
  Casilla 653, Santiago, Chile.\\
 $^2$ Departamento de F\'{\i}sica, Facultad de Ciencia,
Universidad de Santiago de Chile,\\
 Casilla 307, Santiago, Chile. }
\date{ \today }
\maketitle
%%%%%%%%%%%%%%%%%%%%%%%%%%%%%%%%%%%%%%%%%%%%%%%%%%%%%%%%%%%%%%%%%%%
\begin{abstract}
%%%%%%%%%%%%%%%%%%%%%%%%%%%%%%%%%%%%%%%%%%%%%%%%%%%%%%%%%%%%%%%%%%%
Null and timelike geodesics around a 2+1 black hole are determined.
Complete geodesics of both types exist
in the rotating black-hole background, but not in the spinless
case. Upper and lower bounds for the radial size of the orbits
are given in all cases and the possibility of passing from
one black hole exterior spacetime to another is discussed using
the Penrose diagrams. An analysis of particle motions by means
of effective potentials and orbit graphs are also included.
\end{abstract}
\pacs{04.20.Cv, 04.20.Jb, 04.60.Kz, 04.70.Bw.}
%%%%%%%%%%%%%%%%%%%%%%%%%%%%%%%%%%%%%%%%%%%%%%%%%%%%%%%%%%%%%%%%%%%
\section{Introduction}  \label{introd}
%%%%%%%%%%%%%%%%%%%%%%%%%%%%%%%%%%%%%%%%%%%%%%%%%%%%%%%%%%%%%%%%%%%
 The study of timelike and null geodesics is an adequate way to
visualize the main features of a spacetime. The paths of
freely moving particles and photons in the four-dimensional
Schwarzschild metric is the key to understand various important
physical phenomena: planetary motions, gravitational lensing,
radar delay, etc. The three-dimensional black hole solution
was reported in \cite{banados1} (BTZ solution here after). This
solution is a spacetime of constant negative curvature, but
it differs from anti-de Sitter (adS) space in its global
properties; it is obtained from adS space  through
identifications by means of a discrete subgroup of its isometry
group $SO(2,2)$ \cite{banados2}. Here we attempt a
complete review of the motion of massive and massless
test particles in the BTZ black hole.

The first study of geodesics around this black hole was
done in \cite{gamboa}. In order to obtain the equations of
motion for massive particle, these authors use the
Hamilton-Jacobi formalism. However, the simplicity of this lower
dimensional model makes the use of that method unnecessary; the
symmetries of the BTZ solution enables us to derive
directly the equations for timelike and also null
geodesics. Exact solutions can be found
for the timelike geodesic equations, not only for extreme values of
angular momemtum of the black hole as in \cite{gamboa}, but also
for all cases.
Moreover, a different interpretation about the oscillatory
character of radial timelike solution is given using the Penrose
diagram.

This article is organized as follows.
The section \ref{geoeq} is devoted to derive the geodesic
equations. These are easy obtained using the constants of motion
associated to the two Killing vectors and the normalization
condition for the tangent to geodesic curves.

In the section \ref{sol} we first solve the radial equation and
bounds for radial motion are determined. From this, we
conclude that the rotating black hole is a geodesically
complete spacetime. Moreover, the complete geodesics
 allow the  passage from one exterior black hole spacetime to
another. An explanation based on Penrose diagram is given.

We also note that while massive particles always fall
into the event horizon and no stable orbits are possible
\cite{gamboa}, massless particles can escape. This
important issue makes the analogy between this three-dimensional
model and its four-dimensional counterpart more tight:
thermodynamic phenomena (Hawking radiation) as well as
energy extraction from a rotating black hole (Penrose process)
are possible.
 At last, in this section, the  remaining
geodesic equations are integrated and their solutions are
tabulated together with the radial bounds.

It is possible to analyze, in a direct way, the main
geodesic features by defining an {\em effective potential}
\cite{mtw}. We present this analysis in the section \ref{eff}.
In \cite{gamboa}, only one of the two  possible solutions for the
effective potential was discussed. The study of both solutions
is necessary, for instance, to understand the Penrose process
and to determine all the accesible regions for test particles.
An analysis of dragging of inertial frames is also included.

At the last section, we summarize ours results and discuss
briefly the charged black hole.
%%%%%%%%%%%%%%%%%%%%%%%%%%%%%%%%%%%%%%%%%%%%%%%%%%%%%%%%%%%%%%%%%%%
\section{Geodesic equations} \label{geoeq}
%%%%%%%%%%%%%%%%%%%%%%%%%%%%%%%%%%%%%%%%%%%%%%%%%%%%%%%%%%%%%%%%%%%
 The action considered in \cite{banados1} is
\begin{equation}
I={1 \over 2 \pi} \int \sqrt{- g}\, ( R + 2 \ell ^{-2} )\, d^2x \,
dt + B, \label{action}
\end{equation}
where $B$ is a surface term, and the radius of curvature
$\ell = (-\Lambda)^{- 1/2}$ provides the length scale necessary
in order to have a horizon (in $2+1$ dimensions the mass is
dimensionless, and  $\Lambda$ is the cosmological constant).
The Einstein equations are solved by the black hole field
\begin{equation}
ds^2 = -N^2 dt^2 + N^{-2} dr^2 + r^2 ( N^{\phi} dt + d \phi )^2,
\label{ds2}
\end{equation}
where the squared lapse $N^2$ and the angular shift $N^{\phi}$ are
given by
\begin{equation}
N^2  = - M + {r^2 \over \ell^2} + {J^2 \over 4 r^2}
\qquad \mbox{and} \qquad \label{lapse}
N^{\phi} = -{ J \over 2 r^2} \,,
\end{equation}
with   $- \infty < t < \infty$ , $0 < r < \infty$ , and $0 \leq
\phi < 2 \pi$.

The constant of integration $M$ is the conserved charge associated
with asymptotic invariance under time displacements (mass)
and $J$ is that associated with rotational invariance (angular
 momemtum).
The lapse function vanishes for two values of $r$ given by
\begin{equation} \displaystyle
r_{\pm} = \ell M^{\frac{1}{2}} \left [{1 \pm \sqrt
 {1 - ({J \over M \ell})^2 }}  \over 2 \right ]^{\frac{1}{2}},
\label{horz}
\end{equation}
where  $r_+$ is the black-hole horizon ($M >0$ and $|J| \leq M
\ell$).

The surface of infinite redshift  is given by the relation
$\mrms{g}{tt}(r = \mrms{r}{erg})=0$, where  $ \mrms{r}{erg} = \ell
M^{\frac{1}{2}} $. These three values obey the inequality
$r_{-} \,  \leq \, r_{+} \, \leq \, \mrms{r}{erg}$.

The Killing vectors associated to the BTZ metric are two
\cite{banados2},  ${\partial / \partial t}$ and ${\partial /
\partial \phi}$. Thus, the  constants of motion along the
geodesic are
 \begin{equation}
E = -g_{ab} \xi^a u^b = \left [-M + {r^2 \over \ell^2} \right ]
\left ({dt \over d \lambda} \right )  + {J \over 2} \left ({d \phi
\over d \lambda} \right ),
\label{E}
\end{equation}
where $\xi^a = ({\partial / \partial t})^a$ denotes the static
Killing vector, and
\begin{equation}
L = g_{ab} \Phi^a u^b = r^2 \left ({d \phi \over d \lambda}\right)
           - {J \over 2} \left ({dt \over  d \lambda} \right ),
\label{L}
\end{equation}
where $\Phi^a = ({\partial / \partial \phi})^a $ is the rotational
Killing vector and  $u^a = {dx^a / d \lambda}$ is the tangent to
a curve  parameterized by $\lambda$, which is normalized by the
condition
\begin{equation}
u^a u_a = - m^2,  \label{norm} \end{equation}
where $m$ is 1 for timelike geodesic and 0 for null geodesics.

The constant E cannot be interpreted as the local
energy of the particle at infinity since the black-hole field is
not asymptotically flat.

Making the following re-scalings,
$$ \begin{array}{ccccccccc}
\hat{r} & =&{\displaystyle \frac{ r}{\ell \sqrt{M}}},&
\hat{\phi}& =& \phi \sqrt{M},
& \hat{t}& =&{\displaystyle \frac{\sqrt{M}}{\ell}t},\\
 \displaystyle \hat{\lambda}& =&{\displaystyle
\frac{\lambda}{\ell}},&  \hat{E} & = & {\displaystyle
\frac{E}{\sqrt{M}}}, & \hat{L}&  =& {\displaystyle \frac{L}{\ell
\sqrt{M}}},\\ & & & \displaystyle \hat{J}& =&
 {\displaystyle \frac{J}{\ell M} \,,} & & & \end{array} $$
and using  (\ref{E}), (\ref{L}) and (\ref{norm}), we obtain the
geodesic equations (in what follows we will omit the caret on
the variables):
\begin{eqnarray}
  r^2 \dot{r}^2 &=& -m^2(r^4-r^2 +\frac{J^{\,2}}{4})+(E^2-L^2)r^2
                   +L^2-JEL, \label{radial} \\
\dot{\phi} &  =  & \frac{(r^2-1)L +\frac{1}{2}JE}
{(r^2-r_{+}^2)(r^2-r_{-}^2)} \,,  \label{phid} \\
  \dot{t} &  = & \frac{ Er^2-\frac{1}{2} JL}
{(r^2-r_{+}^2)(r^2-r_{-}^2)} \,,    \label{td}
\end{eqnarray}
where dot means \  $ \displaystyle \frac{d}{ d \lambda} $.

Equations (\ref{radial}) to (\ref{td}) describe the motion of
test particles in 2+1 black-hole background. From these equations
we will obtain the orbits for massless and massive particles and
the effective radial potential.

%%%%%%%%%%%%%%%%%%%%%%%%%%%%%%%%%%%%%%%%%%%%%%%%%%%%%%%%%%%%%%%%%%%
\section{ Solutions } \label{sol}
%%%%%%%%%%%%%%%%%%%%%%%%%%%%%%%%%%%%%%%%%%%%%%%%%%%%%%%%%%%%%%%%%%%
The radial equation (\ref{radial}) can be integrated directly.
The solution for timelike geodesics ($m=1$) is
 \begin{equation}
    r^2(\lambda) = \frac{1}{2} \left[ \alpha + \gamma \sin
2(\lambda-\lambda_0) \right], \label{rl}
\end{equation} and for null geodesics ($m=0$),
\begin{equation}
r^2(\lambda) = \left\{ \begin{array}{cl}
 \displaystyle \alpha(\lambda-\lambda_0)^2-\frac{\beta}{\alpha}
& \qquad  \mbox{if $\alpha \neq 0 $ } \\
2\sqrt{\beta}(\lambda-\lambda_0) &  \qquad \mbox{if $\alpha=0$
and $\beta \neq 0$}\\ \mbox{constant} & \qquad  \mbox{if $\alpha=0$ and
$\beta=0$,} \end{array} \right. \label{rln}
\end{equation}
with   $\alpha= E^2-L^2 +m^2$, $ \displaystyle \beta=
L^2-JEL-\frac{m^2J^2}{4}$ and $\gamma=\sqrt{\alpha^2 +4 \beta}$.

It is clear from (\ref{rl}) that there always exist a finite
upper  bound \rms{r}{max} for the radial coordinate, which
verifies $ \mrms{r}{max} > r_+ $. Thus, massive particles can not
escape from the black hole. This inequality can be seen as
follows: replacing $r^2$ by $u+r_+^2$,  (\ref{radial}) reads
\begin{equation}
\frac{\dot{u}^2}{4} = -u^2+B_+u+C_+^2, \label{ru}
\end{equation}
where $$ B_{\pm}=E^2-L^2+m^2(1-2r_{\pm}^2) \, \mbox{ and }
\, \displaystyle C_{\pm}= r_{\pm}\left( E-\frac{JL}{2r_{\pm}^2}
\right).$$
It is easy to see that the  right hand side of (\ref{ru}) has two
real roots for $u$, one positive,  $u_1$, and the other
negative, $-u_2$. Thus, $\dot{u}$ is real only for $ -u_2 \,\leq
\, u \, \leq \,u_1 $, this implies that $\mrms{r^2}{max}=r_+^2 + u_1
\, > \, r_+^2 $ since $u_1$ is positive.
We also note from (\ref{radial}) that there is a lower bound
(\rms{r}{min}) different from zero if
\begin{equation}
 \alpha > 0 \quad \mbox{and} \quad \beta < 0. \label{cond}
\end{equation}
Similarly, replacing $r^2$ by $v+r_-^2$ in Eq.\ (\ref{radial}), we
conclude that the lower bound verifies $\mrms{r}{min} < r_-$.

If the conditions (\ref{cond}) are not satisfied,
timelike geodesics terminate at the singularity $r=0$ \footnote{
The nature of this singularity is discussed in Ref. \cite{banados2} }.
Hence, this spacetime is timelike geodesically complete, {\it
i.e.}, all timelike geodesics, with the exception of those which
terminate at the singularity, have infinite affine lengths both
in the past and future directions. A necessary condition for
$ \beta < 0 $ is $J \neq 0 $, hence all the massive particles hit the
singularity in the spinless black hole.

In the same way, we observe from (\ref{rln}) that this spacetime is
also null geodesically complete provided that $J$ doesn't
vanish.  Hence the rotating solution is geodesically complete.

Unlike the case $m=1$, for null geodesics \rms{r}{max} can be
infinite.  This occurs if $E^2 \geq L^2$ ($\alpha \geq 0$). Thus,
massless particles can escape from the black hole.

Another remarkable feature of the massless case is the
existence of circular orbits of any radius.
This occurs if $\alpha$ and $\beta$ vanish simultaneously which
corresponds to set $E=JL$ with $|J|=1$ (extreme black hole).
The same condition allows a circular timelike geodesic \cite{gamboa},
but its radius is just the horizon.

 In order to solve the geodesic equations (\ref{phid}) and
(\ref{td}), we consider three cases, $J=0$, $0 < |J| < 1$, and
$|J|=1$. The solutions for $\phi$ and $t$ are shown in
Table \ref{table1} and the radial bounds in Table  \ref{table2}.
The trajectories of massless and
massive particles are drawn in the Figs. \ref{o1} and \ref{o2}.

The radial coordinate for timelike geodesic is a periodic
function of the affine parameter $\lambda$
(\ref{rl}). The bounds for this motion are higher
and lower than the outer and inner horizons respectively. One
can  understand this behavior using the Penrose diagram for the
rotating black hole \cite{banados2}, that we include in the Fig.
\ref{pd}. The diagram is formed by an infinite sequence of regions
I $(r_+ \, \leq \, r \,< \, \infty)$, II $( r_- \, < \, r \, <
\, r_+)$ and III $( 0 \, < \, r \, \leq \, r_-)$. We see that the
timelike geodesic $\cal A$ that begin at a point in region I can
cross $r_+$ and $r_-$ and hits the singularity. However, the
geodesic $\cal B$ skips the singularity, crossing $r_+$ and $r_-$
infinitely many times. It is essentially a consequence of extending
maximally the 2+1 spacetime and of the timelike character of the
singularity, as shown in the Fig. \ref{pd}. Notice from the expression
for the $t$ coordinate given in Table \ref{table1}, any radiation
emitted by a particle crossing $r_+$, will be infinitely
red-shifted to an external observer at I, and infinitely
blue-shifted when crossing $r_-$.

%%%%%%%%%%%%%%%%%%%%%%%%%%%%%%%%%%%%%%%%%%%%%%%%%%%%%%%%%%%%%%%%%%%
\section{Effective potential} \label{eff}
%%%%%%%%%%%%%%%%%%%%%%%%%%%%%%%%%%%%%%%%%%%%%%%%%%%%%%%%%%%%%%%%%%%
The radial equation (\ref{radial}) is a quadratic function
of $E$, and can be written as
\begin{equation}
 \dot{r}^2 =(E-\mrms{V}{eff}^+)(E-\mrms{V}{eff}^-),
\end{equation}
where the roots $\mrms{V}{eff}^{\pm}$ are
\begin{equation}
\mrms{V}{eff}^{\pm}(r) = {JL \over 2r^2} \pm \frac{1}{r^2} \sqrt {
 (r^2-r_+^2)(r^2-r_-^2)(L^2  + m^2r^2 )}.
\label{veff}
\end{equation}
Both roots coincide at the event horizon.

It is straightforward, but long, to show that for timelike
geodesics $\mrms{V}{eff}^+$ ($\mrms{V}{eff}^-$) is a monotonically
increasing (decreasing) function in region I; this
implies that there are no stable orbits in that region. We also
see that it is not possible to define the effective potential
in region II. For $m=0$, $\mrms{V}{eff}^\pm$ tends
asymptotically to $\pm L$,
except for the extreme black hole. In that case,
$r_+^2=r_-^2=1/2$, and the effective potential has one constant
branch allowing a circular orbit, as
mentioned in Sec. \ref{sol}.

\subsection{ Non-rotating case }
In Fig. \ref{figeff1} the effective potentials
for $M>0$, $M = 0 , -1$ are shown. The case
$M = -1$ corresponds to an adS space; bound orbits are allowed.
The vacuum state of the black-hole configuration is obtained
for $M = 0$, ($r_+=0$). In this case
all particles fall to the origin.

Fig. \ref{figeff2} shows the effective potentials for null
geodesics. We note that in the cases $M > 0$ massless particles
 with $E\geq L$ can escape from the black hole.
 Moreover, these potentials have a similar form to those for
 massive particles with zero
angular momentum in the four-dimensional Schwarzschild black hole.

\subsection{ Rotating case}
The effective potentials for massive and massless particles are
shown in
figures \ref{figeff3} and \ref{figeff4}. As in the case with $J=0$,
only massless particles can escape from the black hole. When $JL >
0$ only test particles with $E > 0$ have positive local energy
measured in a locally nonrotating frame (LNRF). Nevertheless, when
$JL < 0$ particles with positive energy in the LNRF can have
$E < 0$. These negative energy state exist in the ergoregion
and enables a Penrose process as in the four-dimensional
case \cite{schutz}. Nevertheless, as massive particles can not
 escape from the black hole, energy extraction is only
 possible with massless particles.

When $J$ is different from zero, the inertial frames are
dragged with angular velocity $w$ given by
\begin{equation}
w = {J \over 2 r^2}\,.
\label{wdragg}
\end{equation}
In the four dimensional Kerr black hole the angular velocity of
the frame dragging depend on mass, has the same sign as $J$ and
it falls off for large $r$ as $r^{-3}$.
%%%%%%%%%%%%%%%%%%%%%%%%%%%%%%%%%%%%%%%%%%%%%%%%%%%%%%%%%%%%%%%%%%%
\section{Summary and discussion}  \label{sum}
%%%%%%%%%%%%%%%%%%%%%%%%%%%%%%%%%%%%%%%%%%%%%%%%%%%%%%%%%%%%%%%%%%%
In this article we have solved exactly the timelike  and null
geodesic equations for the 2+1 black hole. The solutions reflect the
singularities of coordinates at the horizons as can be seen from
Table \ref{table1} and Fig. \ref{o2}. We show that only the
rotating  black
hole is geodesically complete. The spacelike character of the
singularity at $r=0$, however, makes the spinless black hole
geodesically incomplete. Since
massive particles cannot escape from black hole, the Penrose
process can take place with massless particles only.

The metric of the charged 2+1 black hole \cite{banados1} is obtained
by  performing the change $N^2 \longrightarrow N^2+\frac{1}{2}Q^2
\log\,(r/r_0)$, where $Q$ is the electric charge of black hole
and $r_0$ is a constant. In this case we can have two, one or no
horizons \cite{achucarro} and then the effective potential could
have local minima and produce a very different behavior for test
particles. Therefore, the charged case requires
further  study in order to reveal its geodesic structure.
%%%%%%%%%%%%%%%%%%%%%%%%%%%%%%%%%%%%%%%%%%%%%%%%%%%%%%%%%%%%%%%%%%%
\acknowledgments
%%%%%%%%%%%%%%%%%%%%%%%%%%%%%%%%%%%%%%%%%%%%%%%%%%%%%%%%%%%%%%%%%%%
Useful discussions with M. Ba\~{n}ados, J. Gamboa, A. Gomberoff,
N. Zamorano
and J. Zanelli are gratefully acknowledged. N. C holds a
Grant No. PG/083/93 of Departamento Postgrado y Post\'{\i}tulo,
Universidad de Chile, C. M thanks a fellowship from CONICYT.
This work was supported in part by Grants Nos. 2930007/93 and
193.0910/93 of FONDECYT (Chile) and Grant No. 04-9331CM of
DICYT, Universidad de Santiago.
%%%%%%%%%%%%%%%%%%%%%%%%%%%%%%%%%%%%%%%%%%%%%%%%%%%%%%%%%%%%%%%%%%%

%%%%%%%%%%%%%%%%%%%%%%%%%%%%%%%%%%%%%%%%%%%%%%%%%%%%%%%%%%%%%%%%%%%
\begin{figure}
\caption{ Penrose diagram for the maximally extended nonextremal
rotating black hole ($ 0 < J^2 < 1 $). } \label{pd}
\end{figure}

\begin{figure}
\caption{ Orbits around the spinless black hole a) $m=1$
 with  $E=2$ and $L=2$, b) $m=0$ with $E>L$ and $E<L$.}
\label{o1}
\end{figure}

\begin{figure}
\caption{ Orbits around the rotating black hole
 a) $m=1$ with $E=2$, $L=2$, $J=0.5$,
b) $m=1$ with $E=2$, $L=2$, $J=-0.5$. As it can be seen from
Table I, the expression that relates $\phi$ and $r$ diverges at
the horizons. Thus, the trajectories approaching the horizons
will spiral round them an infinite number of times.}
\label{o2}
\end{figure}

\begin{figure}
\caption{ Examples of effective potential for massive
particles for different values of $M$.}
\label{figeff1}
\end{figure}

\begin{figure}
\caption{ Examples of effective potential for massless
particles for different values of $M$.}
\label{figeff2}
\end{figure}

\begin{figure}
\caption{ Effective potential for massive
particles with $JL < 0$ and $JL > 0$. The regions between
$\mrms{V}{eff}^+$ and $\mrms{V}{eff}^-$ are forbidden. }
\label{figeff3}
\end{figure}

\begin{figure}
\caption{ Effective potential for massless
particles with $JL < 0$ and $JL > 0$. The regions between
$\mrms{V}{eff}^+$ and $\mrms{V}{eff}^-$ are forbidden. }
\label{figeff4}
\end{figure}
%%%%%%%%%%%%%%%%%%%%%%%%%%%%%%%%%%%%%%%%%%%%%%%%%%%%%%%%%%%%%%%%%%%
%%%%%%%%%%%%%%         ------ Table 1 ------      %%%%%%%%%%%%%%%%%
%%%%%%%%%%%%%%%%%%%%%%%%%%%%%%%%%%%%%%%%%%%%%%%%%%%%%%%%%%%%%%%%%%%
\mediumtext
\begin{table}
\caption{Timelike ($m=1$) and null ($m=0$) geodesics for
different values of $J$. We are  defined
 $ \displaystyle
 f(x\,;B\,;C) =\protect \log
 \left|\protect \frac{x}{2C\protect\sqrt{-m^2x^2+Bx+C^2} +2C^2+Bx}
\right| $ \protect \tablenote{
Notice that $f(x\,;B\,;0)$ is $x$ independent.}
and $ R(x \,;B \,;C)= \protect \sqrt{-m^2x^2+Bx+C^2}. $ }
 \begin{tabular}{cc}
case&orbit equation\\ \tableline
$J=0$
&\begin{tabular}{c} $\displaystyle \phi = \pm \frac{1}{2}
f(r^2\,;E^2-L^2+m^2\,;L) +\phi_0 $ \\
$\displaystyle t = \pm \frac{1}{2}f(r^2-1\,;E^2-L^2-m^2\,;E) +
  t_0$ \\ \end{tabular} \\ & \\
$0 < J^2 < 1$ \tablenote{
The case $J=0$ is obtained directly using
the prescription $ \displaystyle
 \lim_{J \rightarrow 0^{\pm}}\: \frac{J}{r_-} =\:\pm\, 2 $. }
&\begin{tabular}{c}  $\displaystyle \phi = \pm \frac{1}{4}
\frac{J}{\sqrt{1-J^2}}\left[\frac{1}{r_+}f(r^2-r_+^2\,;B_+\,;C_+)-
\frac{1}{r_-}f(r^2-r_-^2\,;B_-\,;C_-) \right] +\phi_0 $
\tablenote{ $B_{\pm}=E^2-L^2+m^2(1-2r_{\pm}^2)$ and $ \displaystyle
 C_{\pm}= r_{\pm}\left( E-\frac{JL}{2r_{\pm}^2} \right)$.} \\
$\displaystyle t  = \pm \frac{1}{\sqrt{1-J^2}} \left[
r_+ f(r^2-r_+^2\,;B_+\,;C_+) - r_- f(r^2-r_-^2\,;B_-\,;C_-)\right]
+t_0 $  \\ \end{tabular} \\ & \\
$J^2=1$ &\begin{tabular}{c} $\displaystyle  \phi  = \mp
\sqrt{\frac{1}{8}}\,J \left[ f(x \,;B \,;C)+ \frac{R(x \,;B \,;C)}
{Cx} \right] +\phi_0$
\tablenote{Here $x=r^2-\frac{1}{2},\: B=E^2-L^2$ and  $C=
\sqrt{\frac{1}{2}}(E-L \,\mbox{sng$J$})$.} \\
$\displaystyle t  = \pm \sqrt{\frac{1}{8}} \left[ f(x \,;B \,;C)-
\frac{R(x \,;B \,;C)}{Cx} \right]+t_0 $  \\ \end{tabular}  \\
 \end{tabular}
\label{table1}
 \end{table}
%%%%%%%%%%%%%%%%%%%%%%%%%%%%%%%%%%%%%%%%%%%%%%%%%%%%%%%%%%%%%%%%%%%
%%%%%%%%%%%%%%%%%%    ------ Table 2 ------        %%%%%%%%%%%%%%%%
%%%%%%%%%%%%%%%%%%%%%%%%%%%%%%%%%%%%%%%%%%%%%%%%%%%%%%%%%%%%%%%%%%%
\begin{table}
\caption{ Radial bounds for timelike ($m=1$)
and null ($m=0$) geodesics for different values of $J$.}
\begin{tabular}{cl}
case&\hspace{4cm} radial bounds\\ \tableline
$J=0$&$\begin{array}{ll}m=1& \qquad  \qquad
 0 \, < \, r^2 \, \leq \, \frac{1}{2} \left [
\sqrt{\left[(E+L)^2+1\right][(E-L)^2 +1]}+E^2-L^2+1  \right ] \\
 m=0& \qquad \qquad
 0 \, < \, r^2 \, \leq \,    \left\{ \begin{array}{cl}
 \infty  & \mbox{if $  E^2 \geq L^2 $}  \\
\displaystyle  \frac{L^2}{L^2-E^2}  & \mbox{if $E^2 < L^2 $}
\end{array} \right.\end{array} $ \\  & \\
$0 < J^2 \leq 1$ &$\begin{array}{ll}m=1& \qquad \qquad
\frac{1}{2} \left [\alpha + \gamma \right]
\tablenote{ $ \alpha= E^2-L^2+1,\; \beta=L^2-JEL-J^2/4, \;
\gamma= \sqrt{[(E+L)^2+1+J][(E-L)^2+1-J]} $. }
 \, \geq \, r^2 \, \geq \, \left\{ \begin{array}{cl}
  \frac{1}{2} \left [\alpha-\gamma \right ]
   & \mbox{if $  \alpha > 0 $ and $ \beta < 0 $}  \\
    0  & \mbox{otherwise}  \end{array} \right. \\
m=0& \qquad \qquad \begin{array}{rcll}   0 \, < \,&r^2& \, <
\infty& \mbox{if $E^2 \ge L^2$ and $L^2 \ge JEL$} \\
 \displaystyle \frac{JEL-L^2}{E^2-L^2} \, \leq \, &r^2& \, <
\infty &  \mbox{if $E^2 > L^2$  and  $L^2 < JEL$ }   \\
0 \, < \,&r^2& \, \leq \displaystyle \frac{L^2-JEL}{L^2-E^2}&\mbox
{if $E^2 < L^2$ and $L^2 > JEL$} \end{array} \end{array}$ \\
 \end{tabular}
 \label{table2}
 \end{table}
%%%%%%%%%%%%%%%%%%%%%%%%%%%%%%%%%%%%%%%%%%%%%%%%%%%%%%%%%%%%%%%%%%%
\end{document}